\documentclass[twocolumn,amsmath,floats,showpacs,nofootinbib]{revtex4}
\usepackage[usenames]{color}
\usepackage{graphicx}
\usepackage{epstopdf}
\usepackage{textcomp}
\usepackage{hyperref}
\usepackage{multirow}
\usepackage{tikz}
\usepackage{rotating}
\hypersetup{colorlinks=true}

\begin{document}

\title{Point-contact spectroscopy of superconducting YBaCuO single crystals of tetragonal modification}

\author{I. K. Yanson, L. F. Rybal'chenko, V. V. Fisun, N. L. Bobrov, V. V. Demirskii, and V. V. Mitkevich}
\affiliation{Physicotechnical Institute of Low Temperatures, Academy of Sciences of the Ukrainian SSR, Kharkov,\\
Email address: bobrov@ilt.kharkov.ua}
\published {(\href{http://fntr.ilt.kharkov.ua/fnt/pdf/14/14-8/f14-0886r.pdf}{Fiz. Nizk. Temp.}, \textbf{14}, 886 (1988)); (Sov. J. Low Temp. Phys., \textbf{14}, 487 (1988)}
\date{\today}

\begin{abstract}The point-contact spectra of tetragonal $\rm YBa_2Cu_3O_{7-\delta}$ with $T_c=50-60~K$ are studied at various temperatures.
The single-electron energy ($2\Delta_0/kT_c\simeq 17$) and its temperature dependence are determined. The characteristic energies of quasiparticle excitations interacting with conduction electrons in this high-$T_c$. superconductor are found to correlate with the characteristic phonon energies. The possibility of a pair gap, which increases with
temperature, is revealed near the critical temperature.


\end{abstract}

\maketitle
Tetragonal crystals of YBaCuO are known to exhibit superconducting properties at $T < T_c \simeq
60~K$ \cite{1}. For pressed point-contacts of the type S-c-N made of this material, we have measured for the first time the current-voltage characteristics (IVC), as well as the firsts ($V_1(eV)\sim dV/dI$) and second ($V_2(eV) \sim d^2V/dI^2$) derivatives of the IVC at different temperatures. A silver or copper electrode N with its edge touching the narrow face of a single crystal YBaCuO (S) was arranged in such a way that the point-contact axis was nearly parallel to the basal plane (see inset to Fig. \ref{Fig1}). Multiple touching and rubbing of the electrodes ensured a metallic conductivity of the junction, i.e., $d^2V/dI^2 > 0$ for $eV\gg\Delta$. This is one of the distinguishing features of our work as compared to the other publications in this field. Superconductivity in the contact region was identified from the appearance of an extremum in the $V_1 (eV)$ dependence, localized at $V = 0$.

Single crystals were grown by spontaneous crystallization from the melt of a nonstoichiometric mixture of initial oxides with molar proportion of metals Y:Ba:Cu = 1:5:14. This method produced crystals with a natural \{100\} facet of size up to 1.5~$mm$ in the plane $ab$ and up to 0.1~$mm$ along the $c$ axis. As in Ref. \cite{1}, we were able to obtain twinless tetragonal crystals of  $\rm YBa_2Cu_3O_{7-\delta}$ which did not assume orthorhombic modification upon a prolonged annealing in an atmosphere of oxygen.

The lattice parameters of the crystals were determined on an automatic diffractometer "Syntex-P2" by the method of least squares from 32 reflexes with $23^{\circ}<\theta<25^{\circ}$ in the $K_{\alpha}$-radiation of Mo. Un- anneaied crystals had the following lattice parameters: $a= 3.868 (1)\text{\AA}$, $c=11.761 (1)\text{\AA}$. Two single crystals (referred to as samples No. 1 and 2 hereafter) were chosen for point-contact studies. The annealing conditions and the lattice parameters of these samples after annealing are presented in Table \ref{tab1}. The criterion for selecting the samples was the perfection of the crystal facets and the existence of a superconducting transition, detected by the inductive method {\renewcommand{\thefootnote}{*}\footnote{The authors thank S.A. Vasil'chenko for inductive measurements.}.

\begin{table}[]
\caption[]{}
\begin{tabular}{|c|c|c|c|c|c|} \hline
{Sample} & \multicolumn{3}{|c|} {Annealing} &\multicolumn{2}{|c|}  {Lattice} \\
\multirow{2}{*}{No} &  \multicolumn{3}{|c|}{parameters} &\multicolumn{2}{|c|} {parameters, \AA} \\ \cline{2-6}
& temp. & time, h & atoms. & \emph{a} & \emph{c} \\ \hline
1 & 650 & 40 & $\rm O_2$ & 3.8641 (9) & 11.718 (3) \\
2& 600 & 5 & Air & 3.8671 (8) & 11.714 (1)\\ \hline
\end{tabular}
\label{tab1}
\end{table}

The contact resistance was found to be $10^{1}-10^{3}~\Omega$, which gives a contact diameter $d\simeq 10^{-5}-10^{-7}~cm$. For such dimensions, the inelastic mean free path of electrons in silver $l_{\varepsilon}(Ag)>d$, and there are no energy losses by the electrons in the current concentration region of the normal electrode. Hence the electrons impinge the superconducting electrode with the maximum surplus energy equal to $eV$.

Figures \ref{Fig1} and \ref{Fig3} show the $V_1(eV)$ characteristics of two different YBaCuO-Ag contacts with a normal state resistance of 20 and 1000~$\Omega$, respectively.
The peak at $V = 0$ indicates the existence of a tunneling component of the current. Although this component is quite small, it allows us to correlate the position of the first $V_1(eV)$ peaks with the energy gap values. For sample No.1 (Fig. \ref{Fig1}),
$\Delta_0\simeq 38~meV$ and $T_c\simeq50\ K$, whence we get
$2\Delta_0/kT_c\simeq 17$. For sample No. 2 (Fig. \ref{Fig3}), it is
difficult to identify the gap singularity. If the first minimum on the $V_1(eV)$ dependence is chosen as the gap singularity, we obtain a value for the gap close to that obtained for Sample No.1. Figure \ref{Fig2} shows the temperature dependence of the gap corresponding to the "indentations" on the $V_1(eV)$ dependence (Fig. \ref{Fig1}). It can be seen that this dependence is lower than the BCS dependence for an isotropic superconductor. Hence, in the high-temperature superconductor (BTC) YBaCuO, as well as in LaSrCuO \cite{2},  the gap is anomalously large and its temperature dependence differs considerably from the BCS dependence due to an anomalously high anisotropy in the physical properties of these layered materials.

\begin{figure}[]
\includegraphics[width=8.5cm,angle=0]{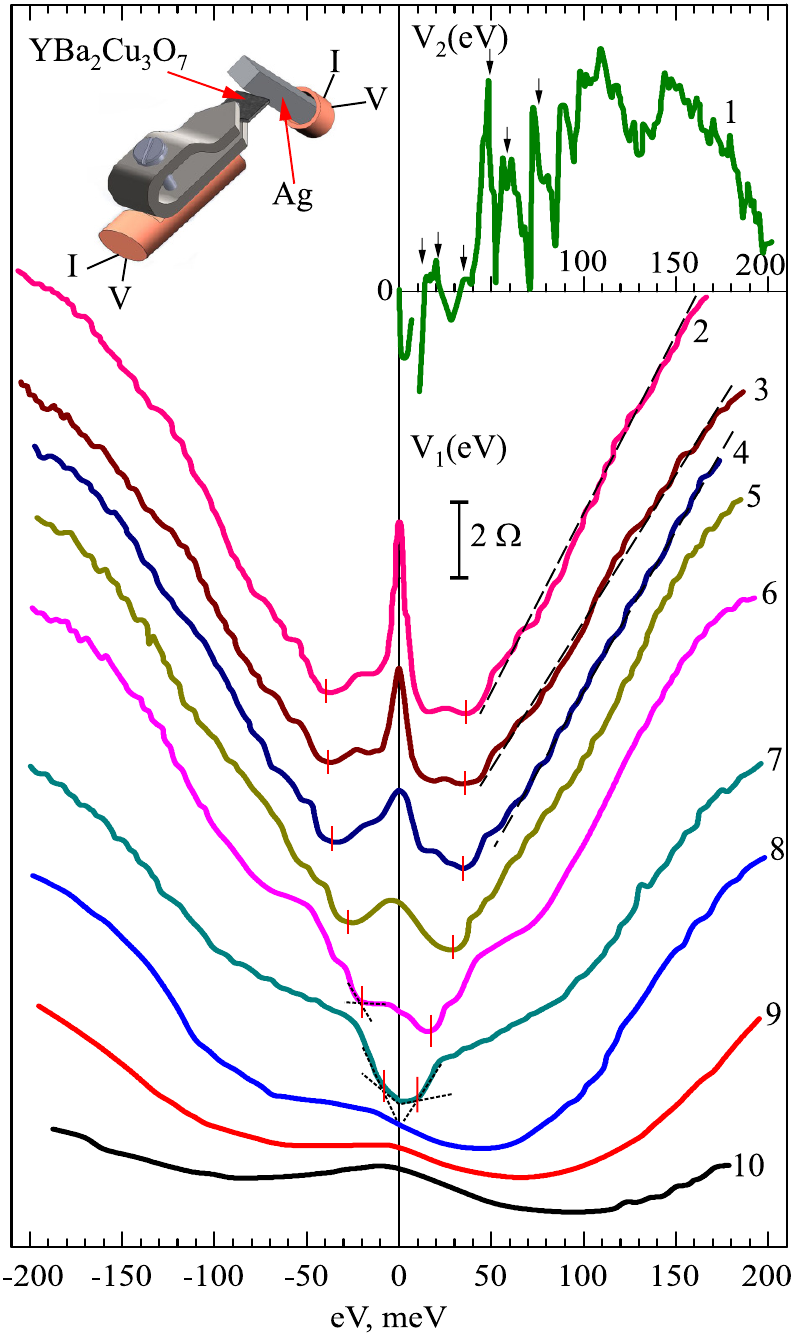}
\caption[]{Dependence of differential resistance ($V_1(eV)\propto dV/dI(eV)$) for sample No.1 at $T$=4.2(1,2); 10(3); 20(4); 30(5); 40(6); 45(7); 50(8); 55(9); 60(10)~$K$. The insets show the geometry of the experiment and the dependence $V_2(eV)\propto d^2V/dI^2(eV)$ at $T=4.2\ K$.}
\label{Fig1}
\end{figure}

\begin{figure}[]
\includegraphics[width=8cm,angle=0]{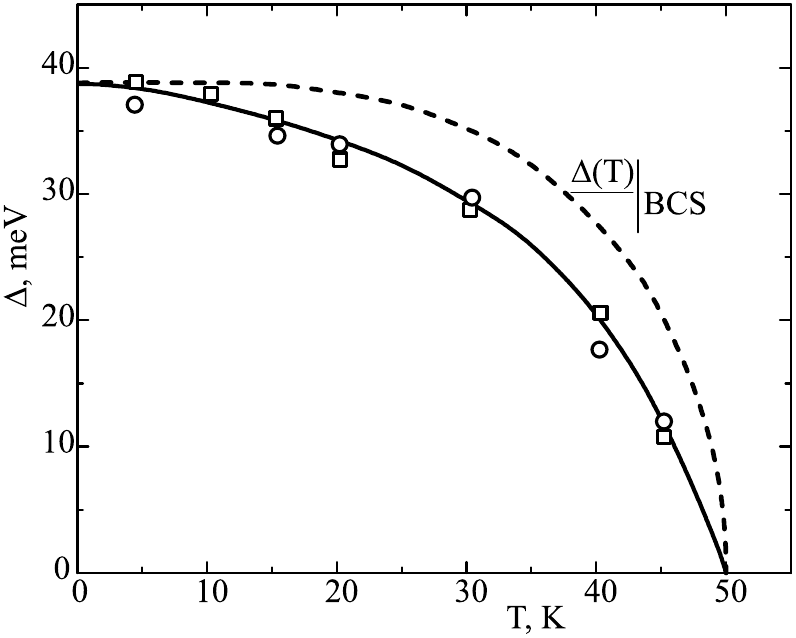}
\caption[]{Temperature dependence of the energy gap. The circles and squares correspond to the gap measurements for positive and negative polarity of $V$ in Fig.\ref{Fig1}.}
\label{Fig2}
\end{figure}

\begin{figure}[]
\includegraphics[width=8cm,angle=0]{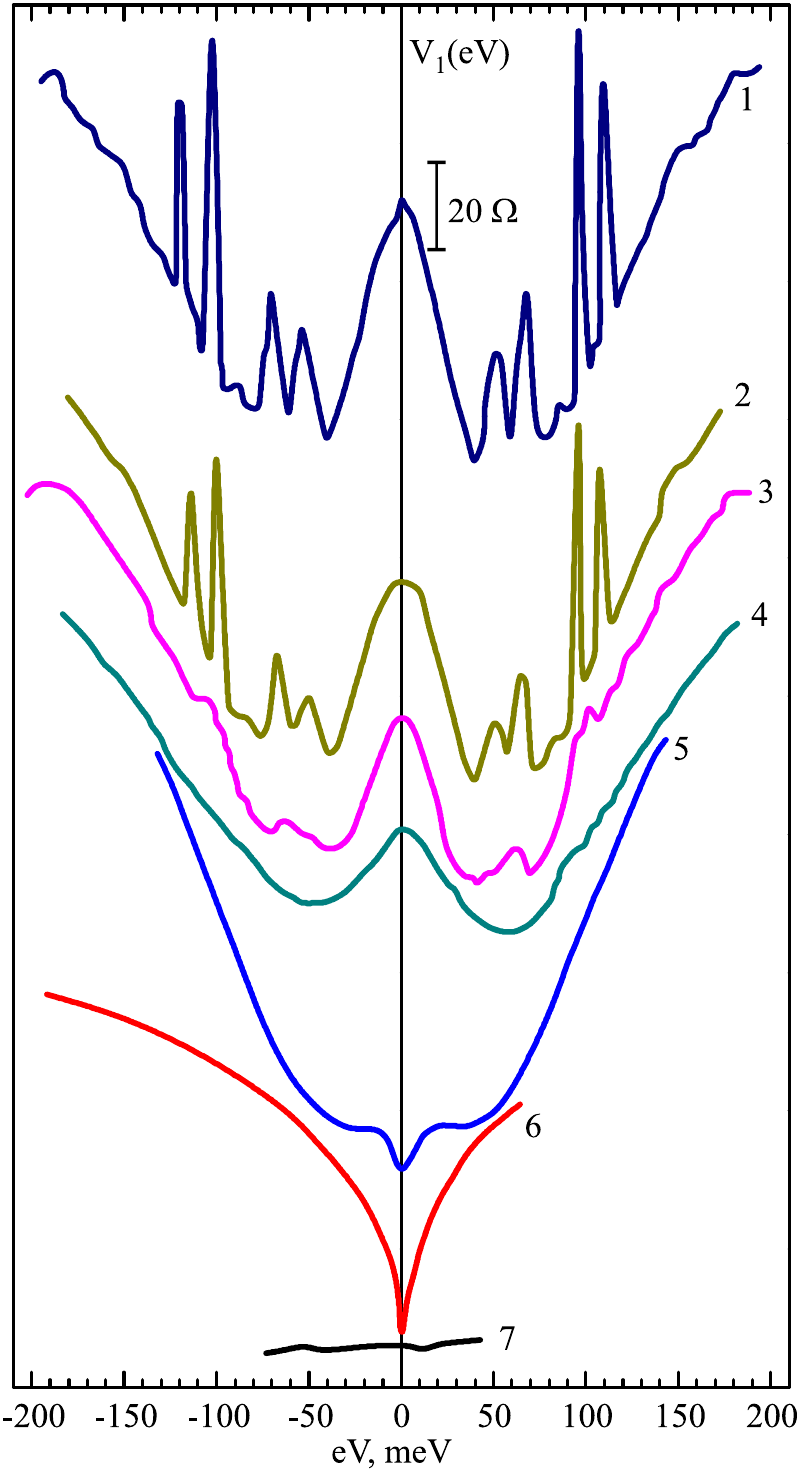}
\caption[]{Point-contact spectra of Sample No.2 for $T$=4.2(1); 10(2); 20(3); 30(4); 40(5); 50(6); 59(7)~$K$.}
\label{Fig3}
\end{figure}

The supplementary structure in the gap region in Fig. \ref{Fig1} and the region beyond the gap in Figs.
\ref{Fig1} and \ref{Fig3} is apparently due to the electron-phonon interaction (EPI). These peculiarities are not associated with superconductivity degradation as a result of heating or to high critical current density since they are practically not displaced on the voltage axis, and their intensity rapidly decreases with temperature. They are too narrow to be
regarded as manifestations of EPI spectrum in IVC, and are obviously caused by nonequilibrium phonons with zero group velocity, which are accumulated near the N-S boundary and which lead to an abrupt decrease in the energy gap of a superconductor in a nonequilibrium state. This causes a decrease in the excess current for displacements equal to characteristic energies of phonons \cite{3}.  A direct spectroscopy of phonons in such superconductors is impossible because of the very small mean free path of charge carriers. As was shown earlier  \cite{3} (see also Ref.
\cite{4}), the nonequilibrium suppression of the gap at point contacts has a tendency to acummulate near the energy of phonons with a low group velocity and a high density of states. Therefore, it is possible to approximately indicate the energies of quasiparticles whose interaction with charge carriers in a superconductor is the strongest. For sample No.
1,	using the second derivative of IVC we obtain
15-22;	36; 46; 57 and 75~$meV${\renewcommand{\thefootnote}{*}\footnote{Other peculiarities are considerably shifted towards lower energies with increasing temperature along the $eV$-axis.}.  For sample No.2, the characteristic energies are obtained from the first derivative (see Fig. \ref{Fig3}): 53; 67; 87; 97 and 112~$meV$. Peculiarities were observed by us at these and close energies for many other contacts with a wide range of resistance.

From the results of neutron diffraction analysis \cite{5},  the phonon spectrum of $\rm YBa_2Cu_3O_7$ terminates at about 90~$meV$. Most of energy values registered by us fit into this interval. High-frequency modes can be due to either phonons in the vicinity of defects (local modes), or quasiparticles other than phonons (which seems to be less probable).

Therefore, the interaction of electrons with local vibrational modes is apparently essential for high-temperature superconductivity in these compounds.

In the superconducting state with a high bias voltage, the differential resistance increases almost linearly with $eV$ in the bias voltage range from 40 to 150~$meV$ (dashed lines in Figs. \ref{Fig1} and \ref{Fig3}), which is similar to the linear temperature dependence of resistivity in the basal plane. This means that the spreading of current in the contact region in the superconducting state takes place predominantly over the basal plane and that potential barriers are absent. Above $T_c$ ($\gtrsim 50~K$ in Fig. \ref{Fig1}), the resistance of contact No.1 at zero bias ($R_0$) radidly increases with temperature, and the IVC curvature gradually changes its sign, acquiring a semiconducting nature. This is apparently due to an increase in the contribution from the direction perpendicular to the basal plane to the contact resistance. The fact that the sign of the curvature $d^2V/dI^2$ of IVC remains positive with increasing $eV$ (see Fig. \ref{Fig1}) and at $T < T_c$ (curve 7) indicates the absence of any noticeable heating by current in the contact region.

A peculiar behavior of $V_1(eV)$-characteristics in the temperature range $T=50-60~K$ corresponding to the superconducting transition for this sample is worth noting (curves 8-10 in Fig. \ref{Fig1}). Broad minima on these curves are of non-heating origin since otherwise the differential resistance should drop rather than increase with voltage at high bias voltages. It is possible that the presence of $V_1(eV)$ in this temperature region corresponds to the energy
gap in the excitation spectrum of local electron pairs existing in these materials at temperatures above $T_c$, as is assumed in some theoretical models \cite{6}.   Moreover, for some combinations of parameters the theory  \cite{7} predicts an increase in the energy gap of pairs as $T_c$ is approached from below, while the "tunnel" gap for quasi-particles vanishes as usual. Figure 1 shows that the "pair gap" observed by us indeed increases as $T_c$ is approached from below.

The $R_0(T)$ dependence for sample No.2 corresponds qualitatively to the standard theory of S-c-N microbridges: as $T\rightarrow T_c$ from below, $R_0$ increases, the rate of increase dropping sharply at $T\gtrsim T_c$. It is noteworthy that the shape of $V_1(eV)$ for this contact changes at $T > 30~K$, i.e., in the region where the gap considerably decreases with temperature, while the coherence length increases. This anomaly is apparently due to a peculiar effect of similarity between HTS and normal metals.

The absence of a broad minimum in the vicinity of $T_c$ in point contact spectra of sample No.2 (see Fig. \ref{Fig3}) can be attributed to a higher resistance of the corresponding contact. As a result, the volume of HTS under investigation is found to be too close to its surface where superconducting parameters are normally suppressed to a considerable extent.

It should be noted in conclusion that singularities of point-contact spectra described in the present paper are not specific for the tetragonal crystallographic modification of $YBa_2Cu_3O_{7-\delta}$. They are also
inherent to a certain extent in the orthorhombic modification with $T_c$ in the vicinity of $90\ K$. The point-contact analysis of the properties of this modification will be published later.

The authors are grateful to B.I. Verkin for his continued interest in this work.

\end{document}